\documentclass{svjour2}  

\usepackage{bbm}
\usepackage{amsmath}
\usepackage{graphicx}
\usepackage{epsfig}

\journalname{Few-Body Syst}

\title{Different Methods for the Two-Nucleon T-Matrix in the Operator Form
}
\author{J.~Golak \and R.~Skibi\'nski \and H.~Wita{\l}a \and K.~Topolnicki \and W.~Gl\"{o}ckle \and A.~Nogga \and H.~Kamada}
\institute{
J.~Golak \and R.~Skibi\'nski \and H.~Wita{\l}a \and K.~Topolnicki 
\at M. Smoluchowski Institute of Physics, Jagiellonian University, PL-30059 Krak\'ow, Poland 
\and W.~Gl\"{o}ckle
\at Institut f\"ur Theoretische Physik II, Ruhr-Universit\"at Bochum, D-44780 Bochum, Germany
\and A.~Nogga 
\at Forschungszentrum J\"ulich, Institut f\"ur Kernphysik, Institute for Advanced Simulation and J\"ulich Center for Hadron Physics, D-52425 J\"ulich, Germany
\and H.~Kamada
\at Department of Physics, Faculty of Engineering, Kyushu Institute of Technology, 1-1 Sensuicho Tobata, Kitakyushu 804-8550, Japan
}
\def\a{\kern+.6ex\lower.42ex\hbox{$\scriptstyle \iota$}\kern-1.20ex a}
\def\e{\kern+.5ex\lower.42ex\hbox{$\scriptstyle \iota$}\kern-1.10ex e}

\begin{document}

\maketitle

\begin{abstract}
We compare three methods to calculate the nucleon-nucleon t-matrix
based on the three-dimensional formulation of Ref.~\cite{2N3DIM}.
In the first place we solve a system of complex linear inhomogeneous equations 
directly for the t-matrix. Our second method is based on iterations
and a variant of the Lanczos algorithm. In the third case 
we obtain the t-matrix in two steps,
solving a system of real linear equations for the k-matrix expansion
coefficients and then solving an on-shell equation, which connects 
the scalar coefficients of the k- and t-matrices. 
A very good agreement among the three methods is demonstrated for
selected nucleon-nucleon scattering observables using a chiral 
next-to-next-to-leading-order neutron-proton potential.
We also apply our three-dimensional framework to the demanding
problem of proton-proton 
scattering, using a corresponding version of the nucleon-nucleon 
potential and supplementing it with the (screened) Coulomb force,
taken also in the three-dimensional form. We show converged
results for two different screening 
functions and find a very good agreement with other methods 
dealing with proton-proton scattering.
\end{abstract}

\section{Introduction}

The nucleon-nucleon (NN) t-matrix appears not only
in the description of the NN scattering process but constitutes 
a key ingredient of the three- and four-nucleon calculations
(see for example \cite{wgrep,Nogga:2000uu}). 
No wonder that many methods to calculate this quantity 
have been proposed
and realized both in coordinate and in momentum space.
For many years it has been natural to use the 
rotational invariance of the NN force and introduce a partial wave basis.
This procedure has a clear physical meaning and is very accurate at low energies.
However, at higher energies very many partial waves
are necessary to achieve converged results, so methods 
employing vector variables might become more efficient.

There have been several approaches to NN scattering avoiding
partial wave decomposition (PWD).
Pioneering calculations of Ref.~\cite{elster98tm} carried out 
for different Malfliet-Tjon-type
potentials gave interesting numerical insight in the properties 
of the three-dimensional (3D) t-matrix
and paved the way for further investigations including spin degrees of freedom.
A helicity formalism directly linked to the total NN spin was
proposed in~\cite{imam}. Later, the spectator equation for relativistic NN scattering 
was solved in~\cite{Ramalho:2006jk} also using a helicity formulation. 
Closely related are 3D formulations for pion-nucleon scattering~\cite{Caia:2003ke} 
and for scattering of protons on light nuclei~\cite{RodriguezGallardo:2008}.

These developments were then used in few-body calculations.
The Faddeev equation for the system of three bosons interacting via scalar forces was 
solved for the bound state case ~\cite{Elster:1998qv} as well as
for three-body scattering~\cite{Liu:2004tv} directly in three dimensions.
The helicity formalism of Ref.~\cite{imam} was employed in 3N calculations 
in~\cite{baye2008,baye2009,baye2011}. 
Recently we proposed a 3D framework
for 3N bound~\cite{2N3N,OUR3NBS3D} and 3N scattering states~\cite{3Nscatt},
where spin-momentum operators are introduced and treated analytically 
so the Faddeev equations turn into a finite set of coupled equations 
for scalar functions depending only on momentum vectors.
For this formulation it is crucial that the most
general form of the NN interaction can only depend on six linearly 
independent spin-momentum operators.
In \cite{2N3DIM} we formulated a new approach to NN scattering 
and provided its numerical
realization based on the matrix inversion method.
A comparison with the standard PWD calculations for two quite different 
NN potentials and a few projectile energies proved 
that the new scheme is very accurate \cite{2N3DIM}.

However, in view of expected applications to the 3N calculations,
where we need the full off-shell t-matrix, we decided to develop 
a more efficient, iteration approach to calculate the 3D t-matrix. 
We also realized an old idea presented, for example, in \cite{book} 
and prepared a scheme for obtaining the on-shell t-matrix from 
the solution of the k-matrix equation.

In Sect.~\ref{section2} we repeat briefly the main points of our approach 
starting from the most general form of the NN potential. 
Next we derive the corresponding
equation for the k-matrix and formulate the subsequent equation
for the on-shell t-matrix. Finally, we formulate our iteration scheme.
Numerical realizations of our different approaches that employ
a recent chiral next-to-next-leading order (NNLO) 
NN force~\cite{Ep05,evgeny.report,newer.report}
are presented in Sect.~\ref{section3}. Here, in addition, we 
apply one of our schemes to the problem of proton-proton 
scattering (including the strong and screened Coulomb forces),
and compare results for two different screening functions.
We conclude in Sect.~\ref{section4}.

\section{The Formal Structure}
\label{section2}

For the convenience of the reader we repeat here the main points of our formalism,
introduced in Ref.~\cite{2N3DIM}.
The NN potential in the two-nucleon (2N) isospin space, spanned 
by the four states
$ \mid t m_t\rangle$
($ \mid 0 0 \rangle$,
$ \mid 1 -1 \rangle$,
$ \mid 1 0 \rangle$,
$ \mid 1 1 \rangle$),
is given as 
\begin{eqnarray}
\langle  t' m_{t'} \mid V \mid t m_t\rangle  = 
  \delta_{ t t'} \delta_{ m_t m_{t'}} V^ { t m_t} \, ,
\label{eq:2.1}
\end{eqnarray}
allowing for charge independence and charge symmetry breaking.
The most general rotational, parity and time reversal invariant 
isospin projected NN force is then expanded into six scalar 
spin-momentum o\-pe\-ra\-tors $w_i ({\vec \sigma}_1,{\vec \sigma}_2, {\bf p'}, {\bf p})$ \cite{wolfenstein} as
\begin{eqnarray}
V^ { t m_t} = \sum_{j=1}^ 6 v_j^ { t m_t} ({\bf p'}, {\bf p}) \;
w_j({\vec \sigma}_1,{\vec \sigma}_2, {\bf p'}, {\bf p}) \, ,
\label{eq:2.3}
\end{eqnarray}
using \cite{2N3DIM}
\begin{eqnarray}
w_1 ( {\vec \sigma}_1,{\vec \sigma}_2, {\bf p'}, {\bf p})&  = &  1 \, , \cr
w_2 ( {\vec \sigma}_1,{\vec \sigma}_2, {\bf p'}, {\bf p})&  = & {\vec \sigma}_1 \cdot {\vec \sigma}_2 \, , \cr
w_3 ( {\vec \sigma}_1,{\vec \sigma}_2, {\bf p'}, {\bf p)}&  = & i \; ( {\vec \sigma}_1
+ {\vec \sigma}_2 ) \cdot ( {\bf p} \times {\bf p'}) \, , \cr
w_4 ( {\vec \sigma}_1,{\vec \sigma}_2, {\bf p'}, {\bf p})&  = & {\vec \sigma}_1
\cdot ( {\bf p} \times {\bf p'}) \; {\vec \sigma}_2 \cdot ( {\bf p} \times {\bf p'}) \, , \cr
w_5 ( {\vec \sigma}_1,{\vec \sigma}_2, {\bf p'}, {\bf p})&  = & {\vec \sigma}_1
\cdot  ({\bf p'} + {\bf p}) \; {\vec \sigma}_2 \cdot  ({\bf p'} + {\bf p}) \, , \cr
w_6 ( {\vec \sigma}_1,{\vec \sigma}_2, {\bf p'}, {\bf p})&  = & {\vec \sigma}_1
\cdot ( {\bf p'} - {\bf p}) \; {\vec \sigma}_2 \cdot  ( {\bf p'} - {\bf p}) \, ,
\label{eq:2.2}
\end{eqnarray}
and scalar functions $v_j^ { t m_t} ({\bf p'}, {\bf p})$ 
which depend only on the ${\bf p}$ and ${\bf p'}$ momenta.
A corresponding expansion into the $w_i ( {\vec \sigma}_1,{\vec \sigma}_2, {\bf p'}, {\bf p})$
operators can be applied to the NN t-operator at the 2N internal energy $z$:
\begin{eqnarray}
t^ { t m_t}(z) = \sum_{j=1}^ 6 t_j^ { t m_t} ( {\bf p'}, {\bf p};z) \;
 w_j( {\vec \sigma}_1,{\vec \sigma}_2, {\bf p'}, {\bf p}) \, .
\label{eq:2.5}
\end{eqnarray}
The $z$ parameter appears in the
Lippmann-Schwinger (LS) equation
\begin{eqnarray}
t^ { t m_t}(z) = V^ { t m_t} + V^ { t m_t} G_0(z) t^ { t m_t}(z),
\label{eq:2.4}
\end{eqnarray}
with $G_0(z)=(z-H_0)^{-1}$ being the free resolvent. 

Inserting Eqs. (\ref{eq:2.3}) and (\ref{eq:2.5})  into the LS equation
(\ref{eq:2.4}), o\-pe\-ra\-ting with  
$w_k( {\vec \sigma}_1,{\vec \sigma}_2, {\bf p'}, {\bf p}) $ from the left and performing the 
trace in the NN spin space leads to
\begin{eqnarray}
\sum_{j} A_{ kj} ( {\bf p'}, {\bf p}) t_j^{t m_t}({\bf p'}, {\bf p};z) = 
\sum_{j} A_{ kj} ( {\bf p'}, {\bf p}) v_j^ { t m_t}( {\bf p'}, {\bf p} )  \; \cr
+ \int \! d^ 3 p'' \sum_{ j j'} v_j^ { t m_t}( {\bf p'}, {\bf p''}) \;
\frac1{z - \frac{{p''}^{\; 2}}{m} + i \epsilon } \;
t_{ j'}^ { t m_t}( {\bf p''}, {\bf p};z) \; B_{ kjj'} ( {\bf p'}, {\bf p''},  {\bf p})
 \, .
\label{eq:2.6}
\end{eqnarray}
The scalar coefficients $A_{kj}$ and $B_{kjj'}$, defined as
\begin{eqnarray}
A_{ kj} ( {\bf p'}, {\bf p}) &\equiv &  {\rm Tr} \Big(w_k( {\vec \sigma}_1,{\vec \sigma}_2, {\bf
p'}, {\bf p}) \;  w_j({\vec \sigma}_1,{\vec \sigma}_2, {\bf p'}, {\bf p})\Big)
\label{eq:2.7} \, , \\
 {B_{ k j j'} ( {\bf p'}, {\bf p''},  {\bf p}) } & \equiv  & 
  {\rm Tr} \Big(w_k( {\vec \sigma}_1,{\vec \sigma}_2, {\bf p'}, {\bf p}) \;
 w_j({\vec \sigma}_1,{\vec \sigma}_2, {\bf p'}, {\bf p''}) \; \nonumber \\ & & \times
  w_{j'}({\vec \sigma}_1,{\vec \sigma}_2, {\bf p''}, {\bf p})\Big) \, ,
\label{eq:2.8}
\end{eqnarray}
are given explicitly in Appendix~A of Ref.~\cite{2N3DIM}.
In this way we obtain a set of six coupled equations for the scalar functions 
$t_j^ { t m_t}({\bf p'}, {\bf p};z)$, which depend, for fixed $z$ and $|{\bf p}|$, on two variables, 
$ | {\bf p'}| $ and the cosine of the relative angle between the
vectors ${\bf p'}$ and ${\bf p}$, given by
${\bf \hat p'} \cdot {\bf \hat p}$.

In Ref.~\cite{2N3DIM} we showed all steps leading to the NN 
elastic scattering observables. The LS equation was solved there 
for positive energy of the NN system,
$z \equiv E_{c.m.} \equiv \frac{p_0^2}{m}$, where $m$ is the nucleon mass.
We chose the following 
representation for the vectors 
${\bf {\hat p}}$, ${\bf {\hat p}'}$ and ${\bf {\hat p}''}$:
\begin{eqnarray}
{\bf {\hat p}} &= &( 0,0,1 ) \, , \nonumber \\
{\bf {\hat p}'} &= &( \sqrt{1 - {x'}^{\, 2}} ,0, x' ) \, , \nonumber \\
{\bf {\hat p}''}& = &( \sqrt{1 - {x''}^{\, 2}} \cos
\varphi'',\sqrt{1 - {x''}^{\, 2}} \sin \varphi'',x'' ) \, ,
\end{eqnarray}
and prepared grids for the
different variables in the problem.
We took $n_x$ Gaussian points for the $x''$ integration and
used the same grid for the $x'$ points.
We employed $n_p$  Gaussian points
for the $p'$ and $p''$ grids, which are defined in the
interval $(0, {\bar p})$. 
In addition, $p_0$ is added to the set of $p'$ points.
The intermediate $\varphi''$ integration was performed 
with $n_{\varphi''}$ Gaussian points.
Thus, from Eq.~(\ref{eq:2.6}) we got a system of $6 \times (n_p+1) \times n_x$ linear equations
for given $p_0$ and fixed $p$. In Ref.~\cite{2N3DIM} we set from the very 
beginning $p=p_0$, so the solution contained the on-shell t-matrix in the operator
form, $t_j(p_0,p_0,x';E_{c.m.})$. Since the six operators (\ref{eq:2.2}) are on-shell 
linearly dependent, our solutions were not unique. They always led, however,
to stable and unique predictions for the observables. In Ref.~\cite{2N3DIM} 
we worked with the standard LU decomposition
of {\it Numerical Recipes} \cite{numrec} for two different NN forces (the Bonn B~\cite{machl}
potential and a chiral NNLO potential~\cite{Ep05,evgeny.report,newer.report}) 
and made calculations for several $E_{c.m.}$ energies.
For all considered cases we obtained a perfect agreement 
between this new 3D approach and the calculations based on
standard partial wave methods.

Solving Eq.~(\ref{eq:2.6}) directly as a system of inhomogeneous coupled algebraic equations
is not efficient. 
Especially in view of the applications in the three-nucleon system, where a full off-shell
t-matrix is needed, it is worth looking for a faster method. 
A typical job employing $n_x = 36 $, $n_p = 36$ 
and $n_{\varphi''}=60$ Gaussian points might take a few hours on a PC. 
In Ref.~\cite{2N3DIM} we tried to solve Eq.~(\ref{eq:2.6}) by iterations.
In the matrix form this equation was written as
\begin{eqnarray}
A t = A v + B t \, ,
\label{eq:mform}
\end{eqnarray}
where $t_j(p',p,x';z)$ ($v_j(p',p,x')$) constituted 
the components of the $t$ ($v$) vector and
\begin{eqnarray}
\left( B t \right)_k (p',p,x') \equiv
\int\limits_0^{\bar p}  dp'' {p''}^{\, 2}
\frac{1}{p_0^2 - {p''}^{\, 2} + i \epsilon } \; f_k ( p''; p',p,x') ,
\label{Sk}
\end{eqnarray}
with
\begin{flalign}
& f_k ( p''; p',p,x') \equiv &
\nonumber \\
& m \sum\limits_{ j, j'=1}^6 \,
\int\limits_{-1}^1 d x'' \int\limits_0^{2 \pi} d\varphi'' \;
 B_{kjj'} ( p', p'', p, x',x'',\varphi'') \; v_j( p', p'',y) \;
t_{ j'}( p'', p, x'' ) \, , &
\end{flalign}
where 
\begin{eqnarray}
y \equiv x' x'' +  \sqrt{1 - {x'}^{\, 2}}  \sqrt{1 - {x''}^{\, 2}} \cos \varphi'' \, .
\label{eq:2.6kkk}
\end{eqnarray}

It was possible to use such $p$, $p'$ and $x'$ points
that the matrix $A$ could be inverted \cite{2N3DIM}. In this case Eq.~(\ref{eq:2.6})
was written as
\begin{eqnarray}
t(p',p,x') = v(p',p,x') + A^{-1}(p',p,x') \;  \left( B t \right) (p',p,x') 
\label{it1}
\end{eqnarray}
so we arrived at the following iteration scheme:
\begin{eqnarray}
 t^{(1)} (p',p,x')& =& v(p',p,x') \, , \nonumber \\
 t^{(n)} (p',p,x')& =& v(p',p,x') \nonumber \\ 
 &+&  A^{-1}(p',p,x') ( B t^{(n-1)} ) (p',p,x') , \ \ {\rm for} \ n > 1 \, .
\label{it2}
\end{eqnarray}
We found it, however, very difficult to maintain numerical
stability for this iterative method. Fortunately, it turned out that 
our scheme could be easily modified to give stable results. 
It is sufficient to combine, on the level of {\em analytical}
calculations, the $A^{-1}_{kj}$ and $B_{kjj'}$ coefficients,
preparing 
\begin{eqnarray}
\tilde{B}_{ kjj'} ( {\bf p'}, {\bf p''},  {\bf p}) \equiv 
\sum\limits_{l=1}^6 A^{-1}_{ kl} ( {\bf p'}, {\bf p}) 
 {B}_{ ljj'} ( {\bf p'}, {\bf p''},  {\bf p}) \, .
\label{Btilde}
\end{eqnarray}
The new, numerically safe, iteration scheme reads thus:
\begin{eqnarray}
 t^{(1)} (p',p,x')& =& v(p',p,x') \nonumber \\
 t^{(n)} (p',p,x')& =& v(p',p,x') +  ( \tilde{B} t^{(n-1)} ) (p',p,x') , \ \ {\rm for} \ n > 1 \, ,
\label{it2n}
\end{eqnarray}
where
\begin{eqnarray}
( \tilde{B} t )_k (p',p,x') \equiv
\int\limits_0^{\bar p}  dp'' {p''}^{\, 2}
\frac{1}{p_0^2 - {p''}^{\, 2} + i \epsilon } \; \tilde{f}_k ( p''; p',p,x') 
\label{Sk2}
\end{eqnarray}
and
\begin{flalign}
& \tilde{f}_k ( p''; p',p,x') \equiv &
\nonumber \\
& m \sum\limits_{ j, j'=1}^6 \,
\int\limits_{-1}^1 d x'' \int\limits_0^{2 \pi} d\varphi'' \;
 \tilde{B}_{kjj'} ( p', p'', p, x',x'',\varphi'') \; v_j( p', p'',y) \;
t_{ j'}( p'', p, x'' ) \, . &
\label{ftilde}
\end{flalign}
The above given iterative method proves also very fast. It is very important 
from the numerical point of view that the summation over $j$ and the integral 
over $ \varphi''$ do not involve $ t_{ j}( p', p, x' ) $, which means that they
can be prepared and stored in advance and later used in each iteration. 
Furthermore, this scheme can be easily implemented on a parallel machine and used to calculate
$ t_{ j}( p', p, x' ) $ for many $p$ points at one shot. 
In the case, when one is interested in the on-shell t-matrix, it is impossible 
to set directly $p=p_0$, if the $p'$ points contain $p_0$. It is sufficient 
to make calculations for two $p$ values, $p=p_0 \pm \delta_{p_0}$ ($\delta_{p_0} \approx 0.01$~fm$^{-1}$)
and then take the average:
\begin{eqnarray}
t_{j}( p_0, p_0, x';z=\frac{p_0^2}{m} ) \approx 
\frac12 \left(
t_{j}( p_0, p_0-\delta_{p_0}, x';z=\frac{p_0^2}{m} ) \right. \nonumber \\
\left.  + t_{j}( p_0, p_0+\delta_{p_0}, x';z=\frac{p_0^2}{m} )
\right) \, .
\label{deltap0}
\end{eqnarray}
This method was successfully employed in the recent 3N bound state 
calculations \cite{OUR3NBS3D}. In this case the kernel contains no pole, which 
makes the calculations even faster. We use a variant of the Lanczos method 
\cite{stadler}
to sum the resulting Neumann series, which proves to be more accurate 
than the standard Pad\'e scheme.

Alternatively, in many calculations using PWD of the NN potential,
the on-shell t-matrix is obtained not directly but 
in two steps. First one defines the k-matrix through the principal 
value kernel \cite{book} 
\begin{flalign}
& \langle {\bf p'} \mid  k^ { t m_t}(z)  \mid {\bf p} \rangle = 
\langle {\bf p'} \mid  V^ { t m_t} \mid {\bf p} \rangle  \nonumber & \\
& + 
\int d^3p'' \;
\langle {\bf p'} \mid  V^ { t m_t} \mid {\bf p''} \rangle \;
\frac{P}{z - \frac{{p''}^{\; 2}}{m} } \;
\langle {\bf p''} \mid  k^ { t m_t}(z) \mid {\bf p} \rangle \, &
\label{kmatrix}
\end{flalign}
and the connection to the t-matrix is given as 
\begin{flalign}
& \langle {\bf p'} \mid  t^ { t m_t}(z)  \mid {\bf p} \rangle =
\langle {\bf p'} \mid  k^ { t m_t}(z)  \mid {\bf p} \rangle \nonumber & \\
& - i \pi \; 
\int d^3p'' \;
\langle {\bf p'} \mid  k^ { t m_t}(z) \mid {\bf p''} \rangle \;
\delta \left( z - \frac{{p''}^{\; 2}}{m} \; \right) \;
\langle {\bf p''} \mid  t^ { t m_t}(z) \mid {\bf p} \rangle \, . &
\label{ktmatrix}
\end{flalign}
If we take $z = \frac{p_0^2}{m}$, 
$ | {\bf p''}|  =  | {\bf p'}|  = | {\bf p}| = p_0$, then 
we get the following (on shell) relation 
\begin{flalign}
& \langle {\bf p'} \mid  t^ { t m_t}(z)  \mid {\bf p} \rangle =
\langle {\bf p'} \mid  k^ { t m_t}(z)  \mid {\bf p} \rangle \nonumber & \\
& - \frac{i}{2} \pi m p_0 \;
\int d {\hat p}'' \;
\langle {\bf p'} \mid  k^ { t m_t}(z) \mid {\bf p''} \rangle \;
\langle {\bf p''} \mid  t^ { t m_t}(z) \mid {\bf p} \rangle \, , &
\label{ktmatrix2}
\end{flalign}
where $ d {\hat p}'' $ denotes the two-dimensional angular integration. 

Including spin degrees of freedom and using the expansion of the k-matrix 
into the $w_i ( {\vec \sigma}_1,{\vec \sigma}_2, {\bf p'}, {\bf p})$
operators (\ref{eq:2.2}), it is easy to find the equation
corresponding to (\ref{kmatrix})
but for the scalar coefficients of the k-matrix, $k_{ j}^ { t m_t}( {\bf p'}, {\bf p};z)$:
\begin{eqnarray}
\sum_{j} A_{ kj} ( {\bf p'}, {\bf p}) k_j^{t m_t}({\bf p'}, {\bf p};z) = 
\sum_{j} A_{ kj} ( {\bf p'}, {\bf p}) v_j^ { t m_t}( {\bf p'}, {\bf p} )  \; \cr
+ \int \! d^ 3 p'' \sum_{ j j'} v_j^ { t m_t}( {\bf p'}, {\bf p''}) \;
\frac{P}{z - \frac{{p''}^{\; 2}}{m} } \;
k_{ j'}^ { t m_t}( {\bf p''}, {\bf p};z) \; B_{ kjj'} ( {\bf p'}, {\bf p''},  {\bf p})
 \, .
\label{eq:2.6k}
\end{eqnarray}
The coefficients $A_{kj}$ and $B_{kjj'}$ are the same as in Eq.~(\ref{eq:2.6}).
By simple interpolation or by taking the average over two $p$ values, 
we get the on-shell k-matrix coefficients, 
$k_{j}^{t m_t}( p_0, p_0, x' ;\frac{p_0^2}{m})$. In order to formulate an equation 
corresponding to (\ref{ktmatrix2}), we have to take into account that on-shell
there are only five independent 
$w_i ( {\vec \sigma}_1,{\vec \sigma}_2, {\bf p'}, {\bf p})$ operators.
This follows directly from the (on shell) relation 
\begin{eqnarray}
{\vec \sigma}_1 \cdot {\vec \sigma}_2 & = &  \frac{ 1}{ ( {\bf p} \times {\bf p'})^ 2 }  \;
{\vec \sigma}_1 \cdot ( {\bf p} \times {\bf p'}) \; {\vec \sigma}_2 \cdot ( {\bf p}
\times {\bf p'})\cr
& + &  \frac{ 1}{ ( {\bf p} + {\bf p'})^ 2 } \; {\vec \sigma}_1 \cdot ( {\bf p} + {\bf
p'}) \; {\vec \sigma}_2 \cdot
( {\bf p} +  {\bf p'})\cr
& + &  \frac{ 1}{ ( {\bf p} - {\bf p'})^ 2 } \; {\vec \sigma}_1 \cdot ( {\bf p} - {\bf
p'}) \; {\vec \sigma}_2 \cdot ( {\bf p} -  {\bf p'}) \, ,
\label{eq:2.10n}
\end{eqnarray}
and can be written as 
\begin{eqnarray}
w_2 = 
\frac1{ p_0^4 \left( 1 - {x'}^{\; 2} \right) } w_4 + 
\frac1{ 2 p_0^2 \left( 1 + {x'} \right) } w_5 + 
\frac1{ 2 p_0^2 \left( 1 - {x'} \right) } w_6 \, .
\label{eq:2.10nn}
\end{eqnarray}
Using Eq.~(\ref{eq:2.10nn}) it is easy to write the on-shell k-matrix 
as a linear combination of only five operators. 
Then the on shell equation corresponding to
(\ref{ktmatrix2}) takes the form 
\begin{flalign}
& \sum\limits_{j=1}^5 A_{ kj} ( p_0, p_0, x' ) t_j^{t m_t}( p_0, p_0,x'; \frac{p_0^2}{m} ) = & \cr
& \sum\limits_{j=1}^5 A_{ kj} ( p_0, p_0, x' ) 
k_j^{t m_t}( p_0, p_0,x'; \frac{p_0^2}{m} ) \; & \cr
& - \frac{i}{2} \pi m p_0 \int\limits_{-1}^1 dx'' \,
\int\limits_{0}^{2 \pi} d{\varphi}'' \, \sum\limits_{j,j'=1}^5 \, 
k_j^{t m_t}( p_0, p_0,y; \frac{p_0^2}{m} ) \; & \cr
& \times \, t_{j'}^{t m_t}( p_0, p_0,x ''; \frac{p_0^2}{m} ) \; 
B_{kjj'} ( p_0, p_0, p_0, x',x'',\varphi'' ) \, . &
\label{eq:2.6kk}
\end{flalign}
In principle, any operator $w_i$ 
appearing in 
(\ref{eq:2.10nn}) can be eliminated but not all the possible 
choices are numerically equivalent due to different behavior of the resulting 
coefficients in the vicinity of $x' = \pm 1$. This is demonstrated 
in Sect.~\ref{section3}. In the case when the operator $w_4$ or $w_5$ is removed,
the $A_{kj}$ and $B_{kjj'}$ coefficients in Eq.~(\ref{eq:2.6kk}) 
have to be renumbered accordingly.

\section{Numerical Results}
\label{section3}

In this section we focus on the on-shell t-matrix 
obtained without employing PWD.
We do not compare our 3D results with the results obtained 
by standard momentum space PWD, since this comparison was 
carried out successfully in \cite{2N3DIM}. 

We show first that the methods discussed in the previous section
lead to stable numerical results. We use the matrix inversion scheme, 
introduced already in \cite{2N3DIM}. 
In present calculations we use a parallel version of the matrix inversion, based on
the SCALAPACK library~\cite{SCALAPACK}.
We implement also the iterative
algorithm and sum up the resulting Neumann series using a
variant of the Lanczos method \cite{stadler}.
Finally, we solve the equation for the k-matrix and, going on-shell,
eliminate either the $w_6$ or the $w_4$ 2N operator. 

We perform calculations using all these methods 
for two projectile laboratory kinetic energies, 13 and 300 MeV.
We use first the neutron-proton version 
of a chiral NNLO NN potential~\cite{Ep05,evgeny.report,newer.report}, 
described briefly in Appendix~A of Ref.~\cite{2N3DIM}. 

In Figs.~\ref{f1} and~\ref{f2} 
we show results for several NN scattering observables.
In the case of 13 MeV, the agreement among the four types of results 
is very good both for the neutron-neutron (left panel) 
and the neutron-proton (right panel) system. 
For the energy of 300 MeV we see that the calculations using the k-matrix 
equation without the $w_6$ operator on-shell, yield slightly different 
results for the $R$ and $A$ observables in the case of the neutron-proton system,
than the other three schemes. These numerical instabilities become better 
visible for the Wolfenstein parameters \cite{book} $c$, $g$, $h$ and $m$ shown in Fig.~\ref{f3}.
Thus this particular numerical scheme should be avoided. 

\begin{figure}[t]\centering
\epsfig{file=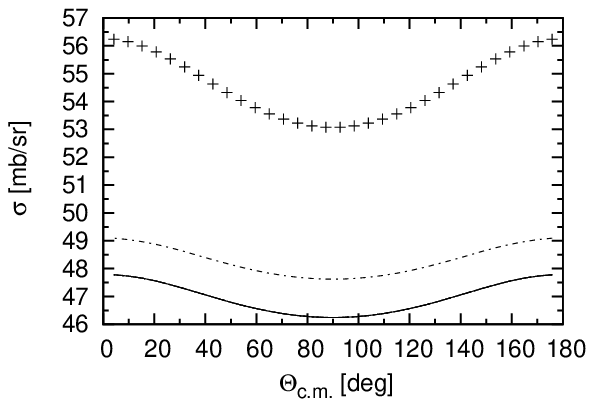,width=5cm,angle=0}
\epsfig{file=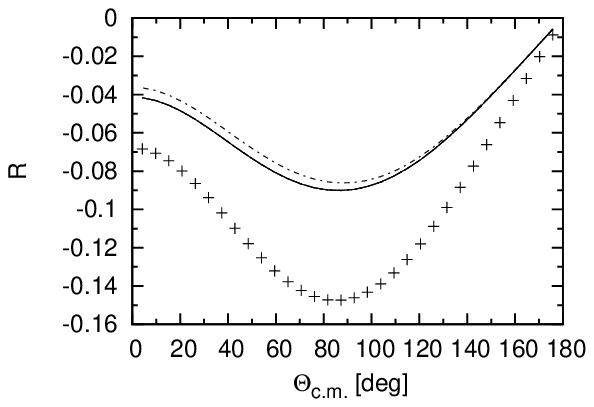,width=5cm,angle=0}
\epsfig{file=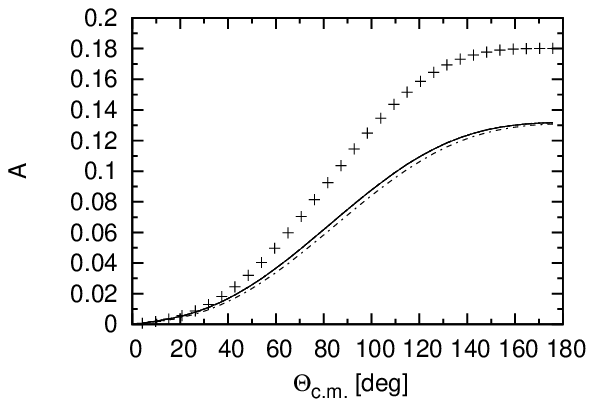,width=5cm,angle=0}
\epsfig{file=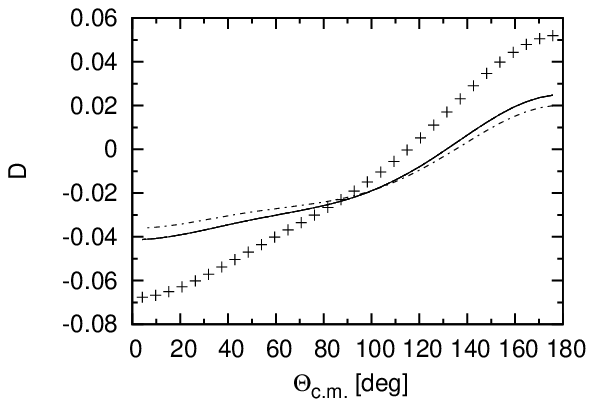,width=5cm,angle=0}
\caption{The convergence of results obtained with the iterative 
method using different numbers
of iterations for neutron-neutron (left panel) 
and neutron-proton (right panel) scattering observables 
for the projectile laboratory kinetic energy
of 13~MeV and the chiral NNLO potential~\cite{evgeny.report}. 
Crosses show the predictions with 6 iterations. Dash-dotted,
double-dotted, dotted, dashed and solid lines display the results 
calculated with 8, 10, 12, 14 and 20 iterations, respectively.
The last three lines overlap.
}
\label{f4}
\end{figure}

\begin{figure}[t]\centering
\epsfig{file=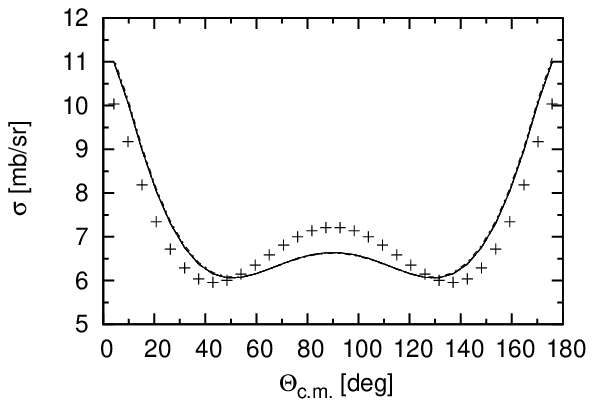,width=5cm,angle=0}
\epsfig{file=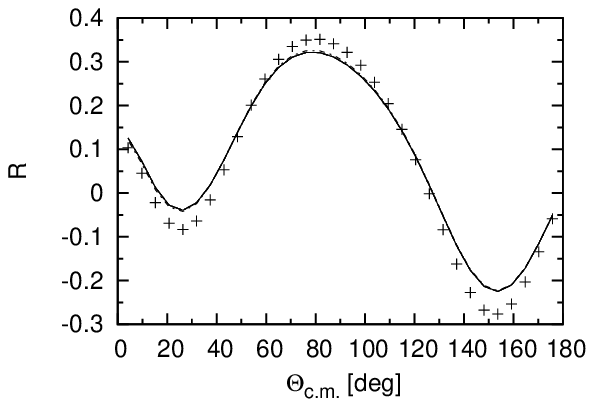,width=5cm,angle=0}
\epsfig{file=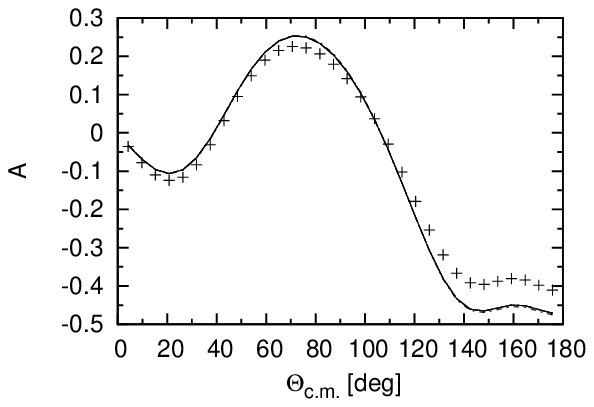,width=5cm,angle=0}
\epsfig{file=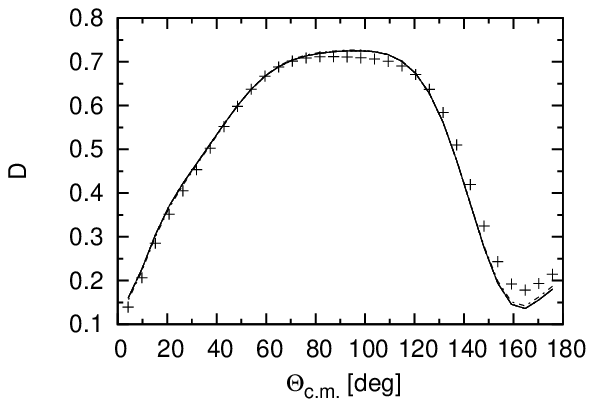,width=5cm,angle=0}
\caption{The same as in Fig.~\ref{f4} for the projectile laboratory kinetic energy
of 300~MeV.}
\label{f5}
\end{figure}

In the case, where iterations are used, we wanted to know how 
many iterations are needed to achieve full convergence
of our results. We expected that the convergence depends on the NN system internal
energy and show in Figs~\ref{f4} and~\ref{f5} our results for various numbers of iterations 
again for the small (13 MeV) and large (300 MeV) laboratory energy. 
At 13 MeV and for the neutron-neutron system we find that 
10 iterations are fully sufficient.
For some observables results obtained with 8 iterations
are very close to the predictions based on 20 iterations.
For the neutron-proton case one needs generally 12 iterations.
At 300 MeV we see even faster convergence.
For the neutron-neutron case predictions based on 6 iterations have already a correct shape
(this was generally not the case for 13 MeV)
and 8 iterations bring us very close to the fully converged results,
although some tiny deviations are still visible. 
For the neutron-proton system results obtained from 10 iterations are still slightly 
different from the fully converged predictions.
We can thus state that for energies below the pion production threshold,
one needs no more than 12 iterations to arrive at fully converged 
results. Since the essential part of the iteration kernel can be prepared in advance, 
iterations run very fast, especially for the case of negative 2N internal energy, 
where no singularity is present. 
This method is considered by us as the best for 3N calculations.

In the remaining part of this section we consider the most difficult 
case in the nucleon-nucleon scattering, namely the proton-proton case with inclusion 
of the Coulomb force. As described in Ref.~\cite{romek}, contrary to the approaches 
using the so-called renormalization procedure (see for example \cite{RodriguezGallardo:2008}),
we add a screened Coulomb potential to the (strong) proton-proton 2N force given
in the operator form and use this sum as input in our 3D calculations. 
We work with the proton-proton version of the same chiral NNLO 2N potential
and choose a relatively small laboratory energy (13 MeV), where Coulomb effects 
are expected to play an important role. 
We assent to two types of screening,
the exponential screening, $s_1 (r;n,R)$,
\begin{eqnarray}
 s_1(r;n,R) \equiv \exp \left( -\left(  
            \frac{r}{R} 
\right)^n \right)  \, ,
\label{s1} 
\end{eqnarray} 
restricting ourselves to the case $n=4$ 
and the localized (transition from $1$ to $0$ takes place
in a finite interval) screening, $s_2 (r;R)$, 
proposed in \cite{RodriguezGallardo:2008}:
\begin{eqnarray}
 s_2(r;R) \equiv 
\theta ({R}-r)+\frac{1}{2} \theta
   (r-{R}) \theta (3
   {R}-r) \left(\sin
   \left(\frac{\pi  r}{2
   {R}}\right)+1\right) \, ,
\label{s2}
\end{eqnarray}  
with $ \theta (x) $ being the (unit) step function. The 3D momentum space 
screened Coulomb potential is obtained as in Ref.~\cite{romek2}.

\begin{figure}[t]\centering
\epsfig{file=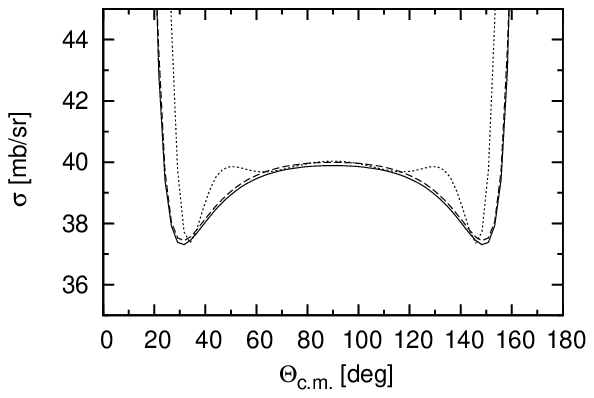,width=5cm,angle=0}
\epsfig{file=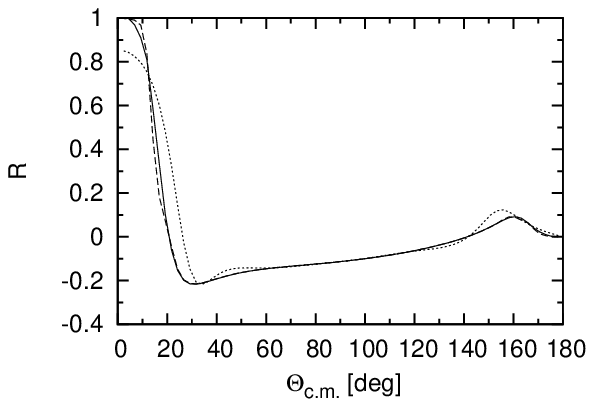,width=5cm,angle=0}
\epsfig{file=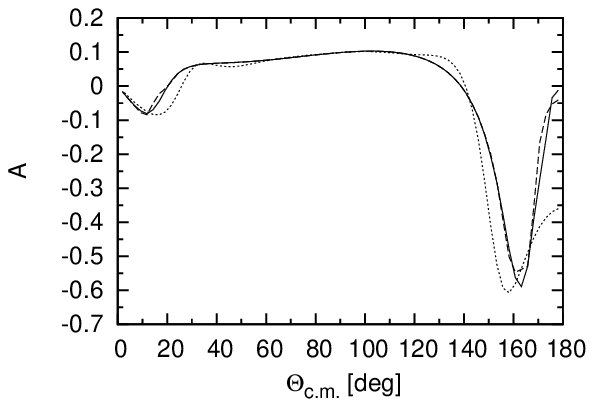,width=5cm,angle=0}
\epsfig{file=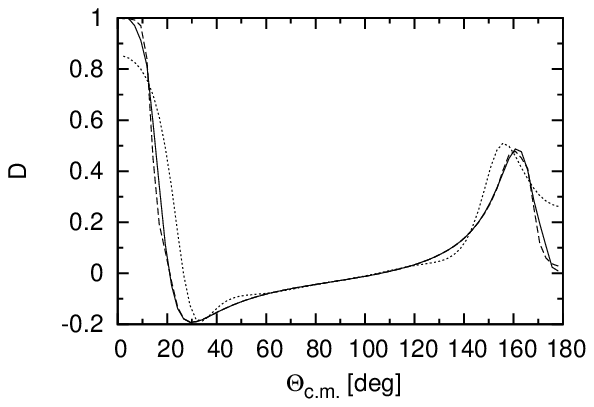,width=5cm,angle=0}
\caption{Selected observables for the proton-proton 
system (including the Coulomb force) at the projectile laboratory kinetic energy 13~MeV
as a function of the center-of-mass angle $\theta_{c.m.}$
for the chiral proton-proton NNLO potential~\cite{evgeny.report,evgenypp}. Results are obtained 
by solving the LS equation using the matrix inversion method
and applying the exponential screening, $s_1(r;n,R)$, to the Coulomb. 
Dotted, dashed and solid lines show the results with the screening parameter $R$= 20, 60 and 120 fm,
respectively, and $n=4$.}
\label{f6}
\end{figure}

\begin{figure}[t]\centering
\epsfig{file=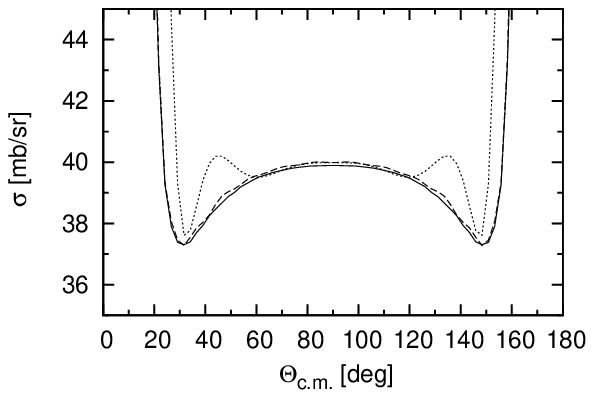,width=5cm,angle=0}
\epsfig{file=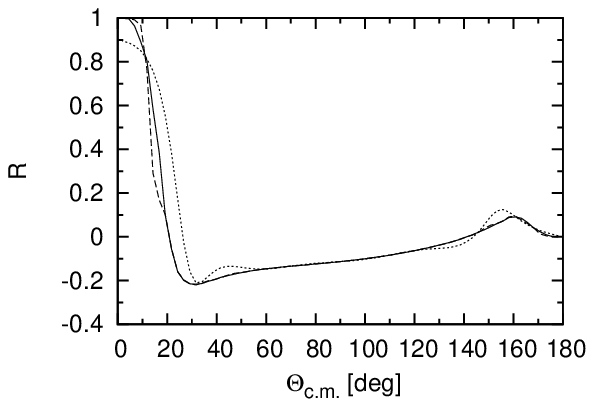,width=5cm,angle=0}
\epsfig{file=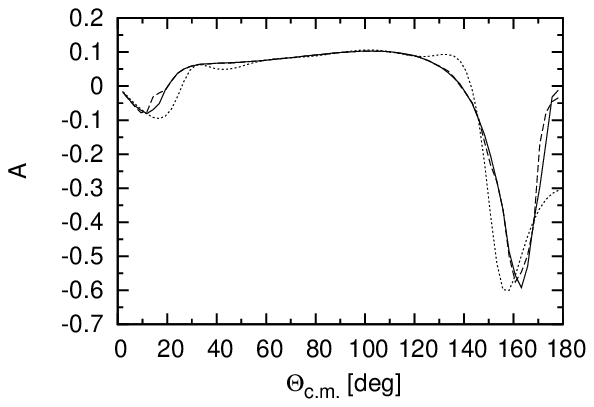,width=5cm,angle=0}
\epsfig{file=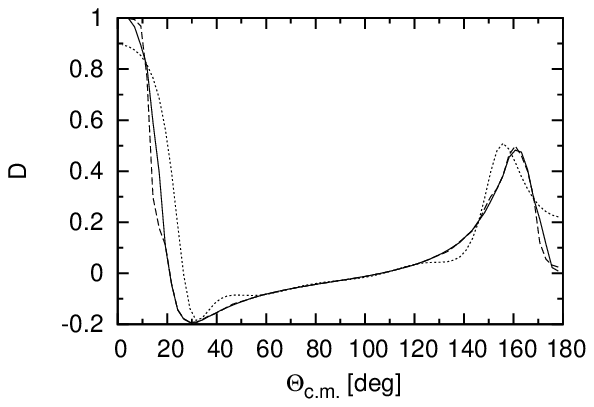,width=5cm,angle=0}
\caption{The same as in Fig.~\ref{f6} with the localized (sine function) 
screening (\ref{s2}), applied to the Coulomb force.
Dotted, dashed and solid lines show the results with the screening parameter $R$= 10, 30 and 55 fm,
respectively.}
\label{f7}
\end{figure}

\begin{figure}[t]\centering
\epsfig{file=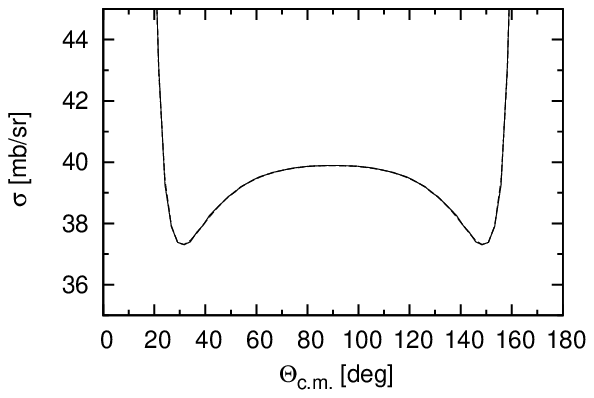,width=5cm,angle=0}
\epsfig{file=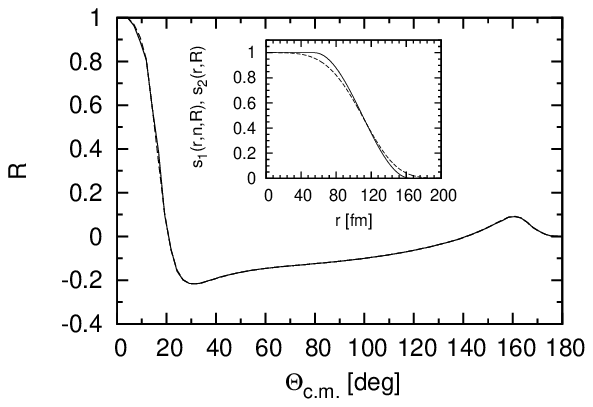,width=5cm,angle=0}
\epsfig{file=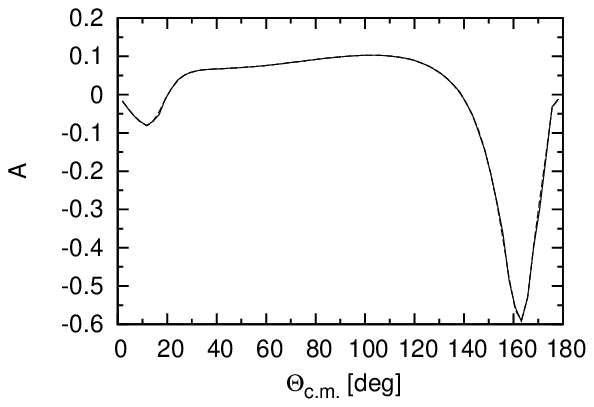,width=5cm,angle=0}
\epsfig{file=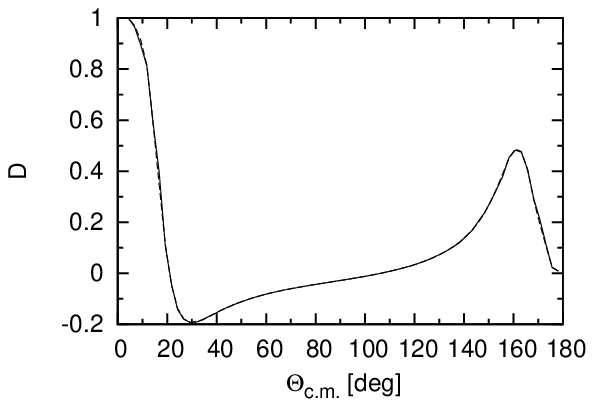,width=5cm,angle=0}
\caption{Selected observables for the proton-proton 
system (including the Coulomb force) at the projectile laboratory kinetic energy 13~MeV
as a function of the center-of-mass angle $\theta_{c.m.}$
for the chiral proton-proton NNLO potential~\cite{evgeny.report,evgenypp}. 
Results with the $s_1(r;n=4,R=120~{\rm fm})$ screening (dashed lines)
are compared with the predictions based on the $s_2(r;R=55~{\rm fm})$ screening
(solid lines).
In the inset the two screening functions are compared: the dashed (solid) line 
shows the $s_1$ ($s_2$) function.
}
\label{f8}
\end{figure}

In Figs.~\ref{f6} and~\ref{f7} we demonstrate a convergence 
of the results for some observables with the increasing parameter $R$. In the case 
of the exponential screening with $n=4$ it is achieved 
for $R$= 120 fm, while for the second type of screening 
$R$= 55 fm is sufficient. In Fig.~\ref{f8} we show that, 
as expected, the two screening prescriptions lead to the same 
final result for the selected proton-proton observables
in our fully 3D calculations. The inset in Fig.~\ref{f8} 
serves to compare the two screening functions 
for the parameters used in the actual calculations. 

\begin{figure}[t]\centering
\epsfig{file=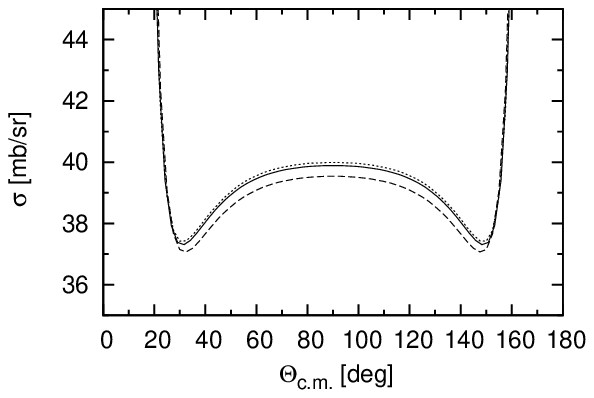,width=5cm,angle=0}
\epsfig{file=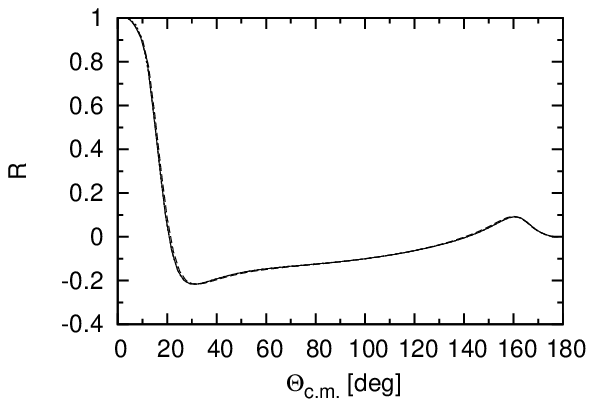,width=5cm,angle=0}
\epsfig{file=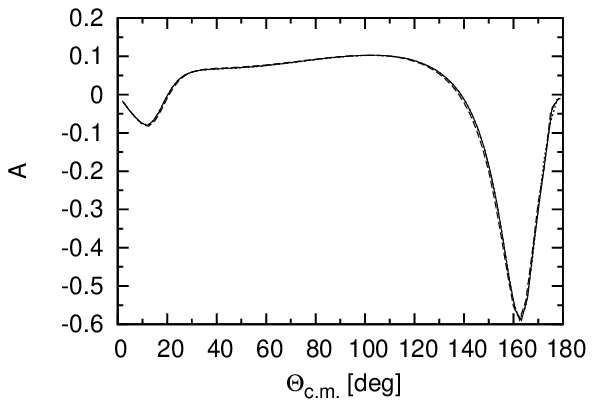,width=5cm,angle=0}
\epsfig{file=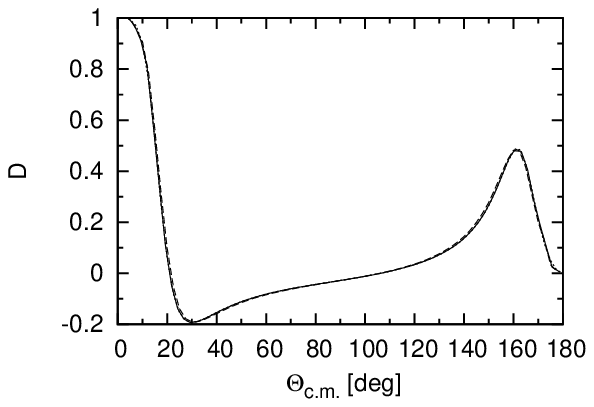,width=5cm,angle=0}
\caption{Selected observables for the proton-proton 
system (including the Coulomb force) at the projectile laboratory kinetic energy 13~MeV
as a function of the center-of-mass angle $\theta_{c.m.}$
for the chiral proton-proton NNLO potential~\cite{evgeny.report,evgenypp}. 
Fully 3D results with the $s_1(r;n=4,R=120~{\rm fm})$ screening (solid lines)
are compared with the predictions based on two methods used in Ref.~\cite{romek}: 
the method combining a 3D Coulomb matrix (for the same screening $s_1$)
with PWD calculations (dashed lines) and
the Vincent-Phatak method \cite{vincentphatak} (dotted lines).}
\label{f9}
\end{figure}

\begin{figure}[t]\centering
\epsfig{file=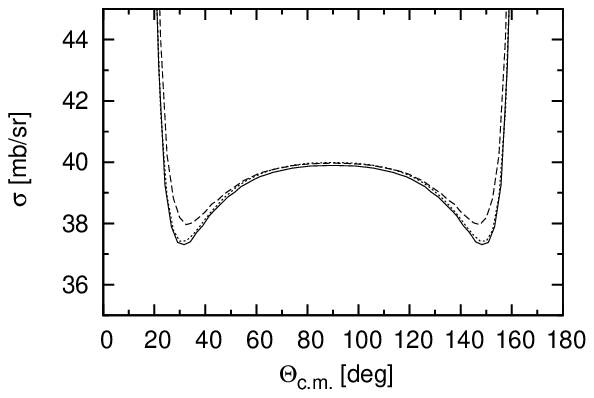,width=5cm,angle=0}
\epsfig{file=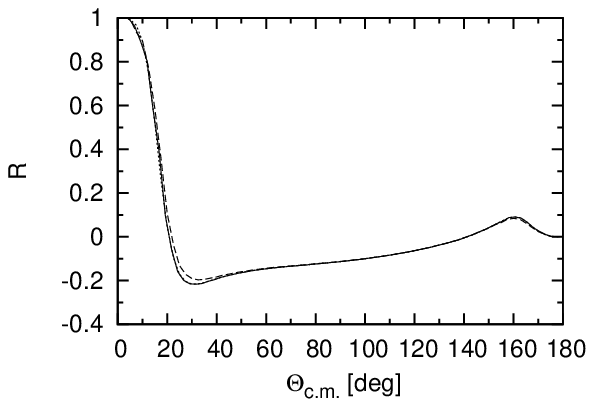,width=5cm,angle=0}
\epsfig{file=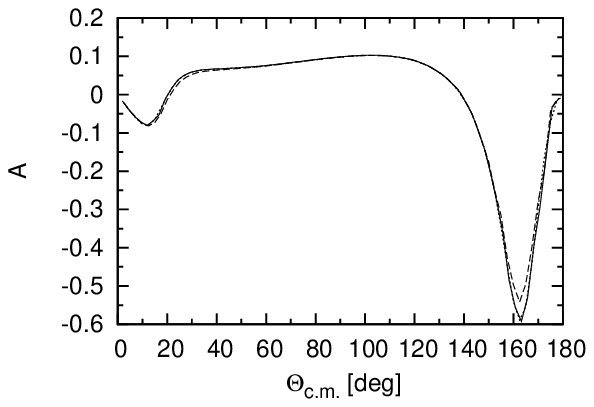,width=5cm,angle=0}
\epsfig{file=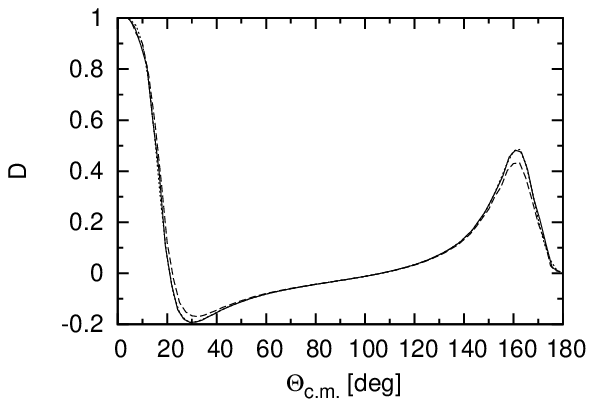,width=5cm,angle=0}
\caption{The same as in Fig.~\ref{f9} but the $s_2(r;R=55~{\rm fm})$ screening 
\cite{RodriguezGallardo:2008} is used in the 3D calculations involving the Coulomb force.}
\label{f10}
\end{figure}

Proton-proton scattering was already considered by us in \cite{romek}. 
There we used the Vincent-Phatak method \cite{vincentphatak} 
and the method combining PWD calculations with the full 3D t-matrix 
of the screened Coulomb potential. Using those two methods,
we performed calculations 
for the proton-proton strong potential and screening functions used in this paper.
Predictions for the $R$ parameters yielding converged results in the case of the fully 3D 
calculations (solid line) are compared in Figs.\ref{f9} (exponential screening) 
and~\ref{f10} (localized screening) with the Vincent-Phatak method (dotted line)
and the method combining PWD and 3D approaches (dashed line).
We observe a good or even a very good agreement among all the methods
for the polarization observables, especially for the exponential screening.
Discrepancies in all the cases, including the most visible ones for the cross section,
do not exceed 1-2~\%. This agreement provides yet another proof that our numerical
treatment of nucleon-nucleon scattering is fully reliable.

\begin{figure}[t]\centering
\epsfig{file=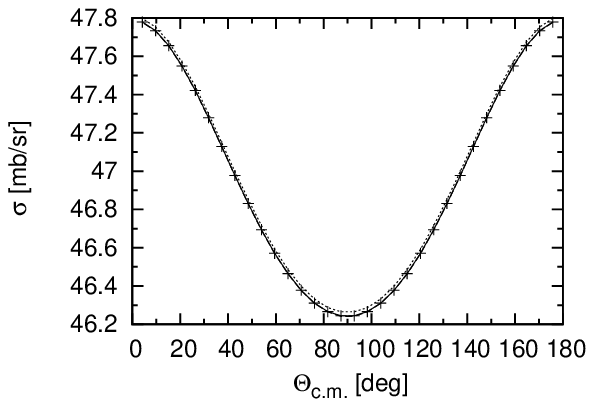,width=5cm,angle=0}
\epsfig{file=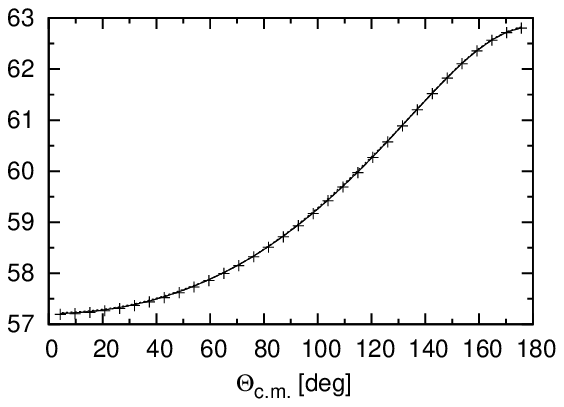,width=5cm,angle=0}
\epsfig{file=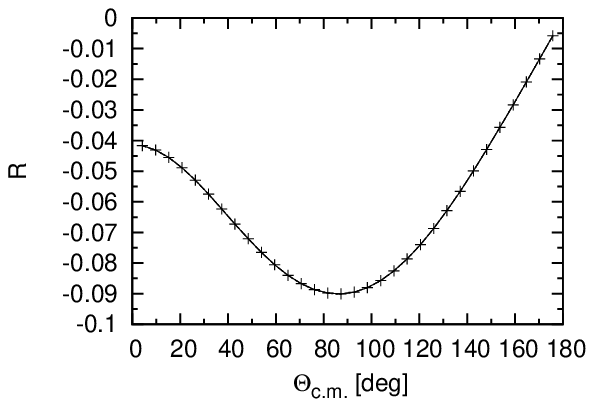,width=5cm,angle=0}
\epsfig{file=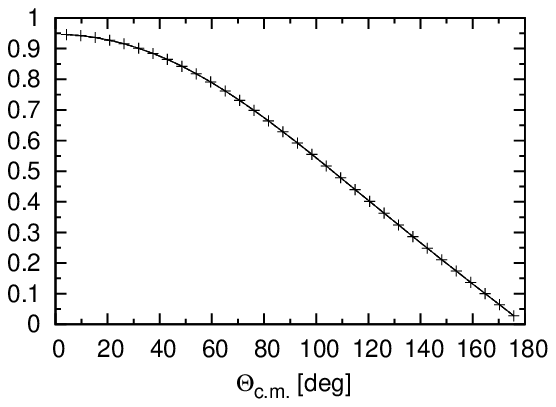,width=5cm,angle=0}
\epsfig{file=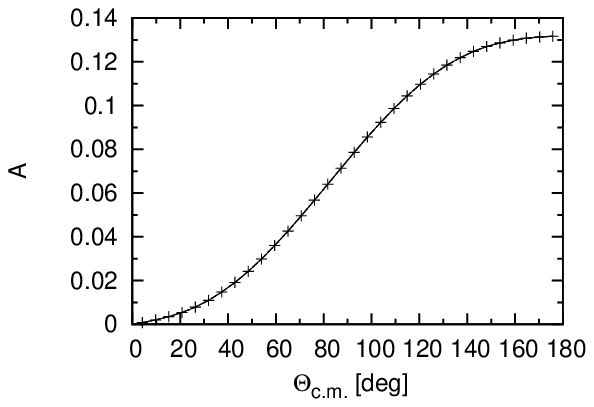,width=5cm,angle=0}
\epsfig{file=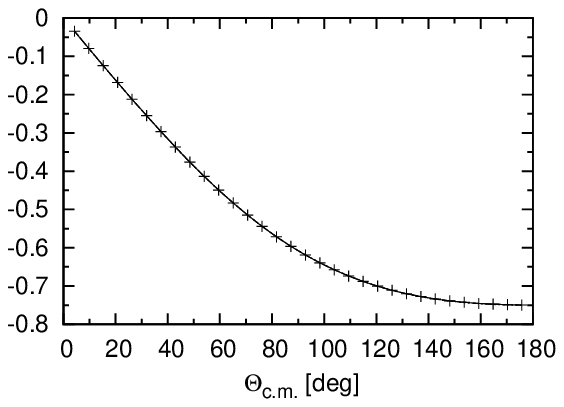,width=5cm,angle=0}
\epsfig{file=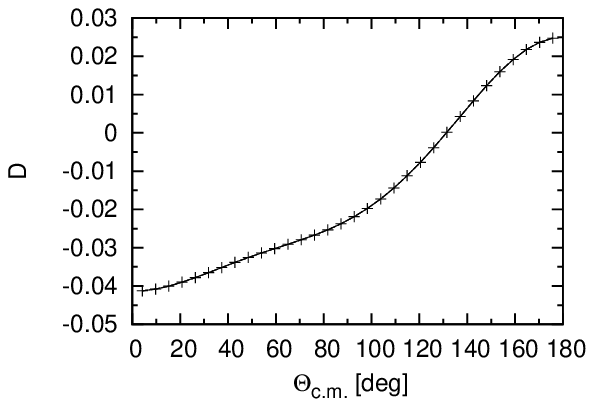,width=5cm,angle=0}
\epsfig{file=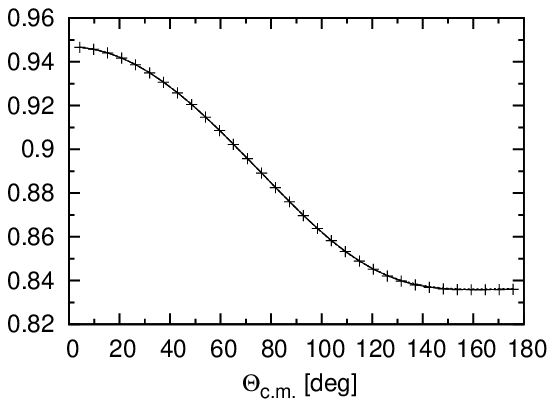,width=5cm,angle=0}
\caption{Selected observables for the neutron-neutron (left panel) 
and neutron-proton (right panel) system at the projectile laboratory kinetic energy 13~MeV
as a function of the center-of-mass angle $\theta_{c.m.}$
for the chiral NNLO potential~\cite{evgeny.report}. Crosses 
represent results obtained with the iterative method (see text).
Dashed (dotted) lines depict the results obtained by solving first the equation 
for the k-matrix and then the on-shell equation for the t-matrix, where on shell 
the $w_4$ ($w_6$) 2N operator is eliminated. Solid lines are for the predictions 
based on the matrix inversion method.
For the definition of the $R$, $A$ and $D$ observables see e.g.~\cite{book}.}
\label{f1}
\end{figure}

\begin{figure}[t]\centering
\epsfig{file=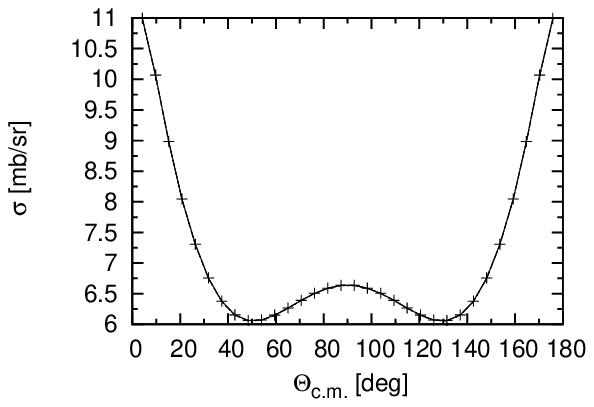,width=5cm,angle=0}
\epsfig{file=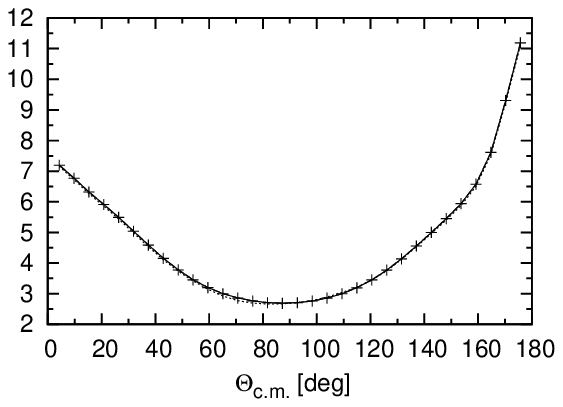,width=5cm,angle=0}
\epsfig{file=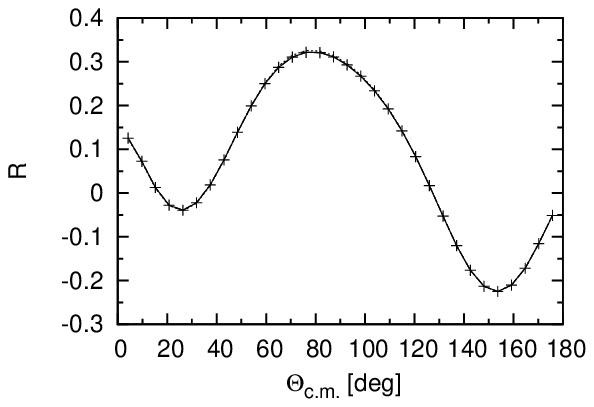,width=5cm,angle=0}
\epsfig{file=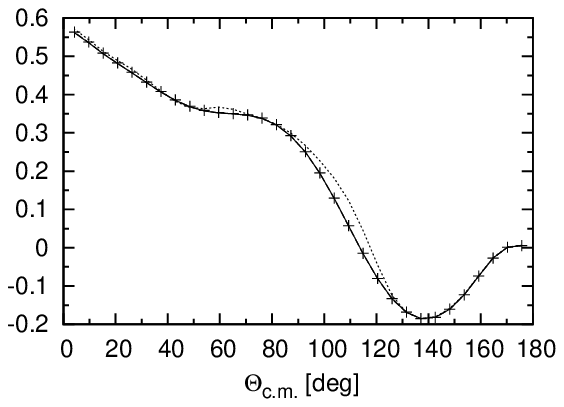,width=5cm,angle=0}
\epsfig{file=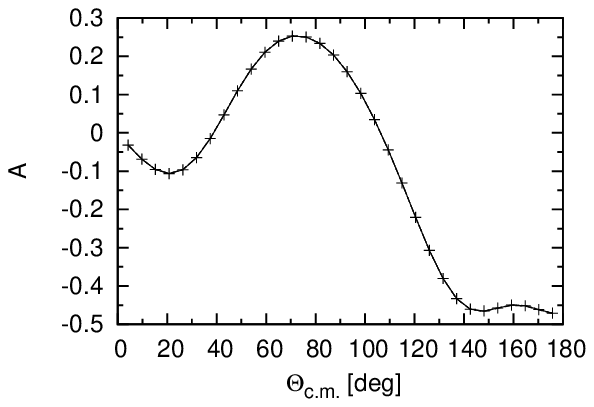,width=5cm,angle=0}
\epsfig{file=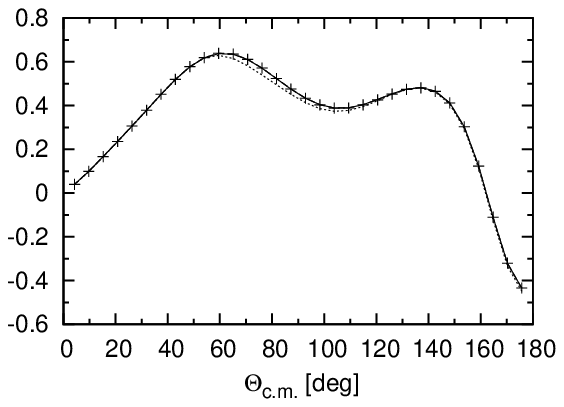,width=5cm,angle=0}
\epsfig{file=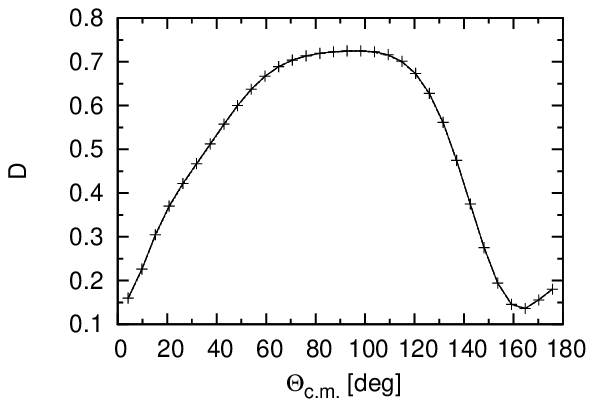,width=5cm,angle=0}
\epsfig{file=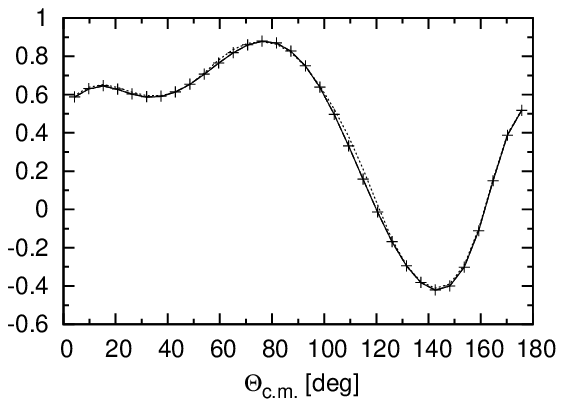,width=5cm,angle=0}
\caption{The same as in Fig.~\ref{f1} for the projectile laboratory kinetic energy
$E$=300~MeV.}
\label{f2}
\end{figure}

\begin{figure}[t]\centering
\epsfig{file=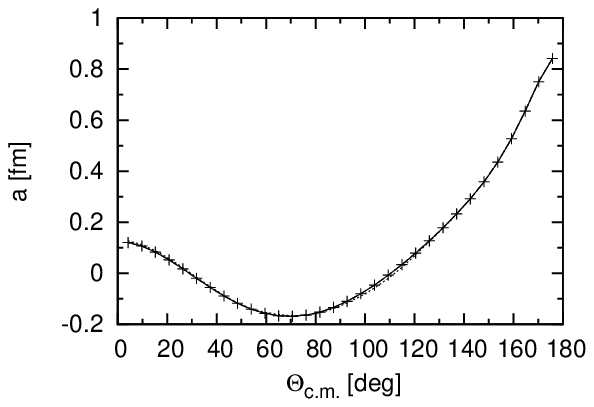,width=5cm,angle=0}
\epsfig{file=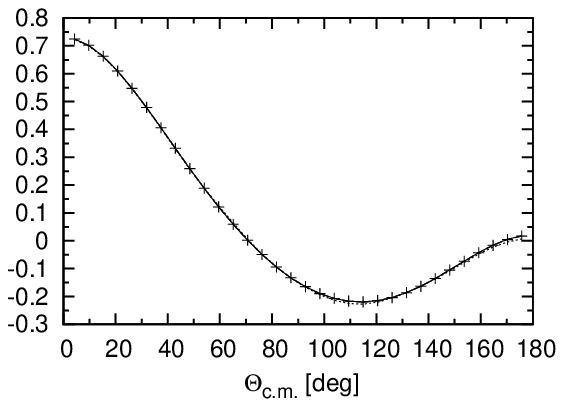,width=5cm,angle=0}
\epsfig{file=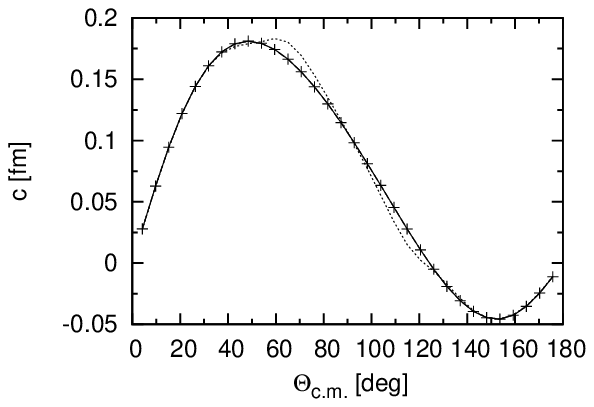,width=5cm,angle=0}
\epsfig{file=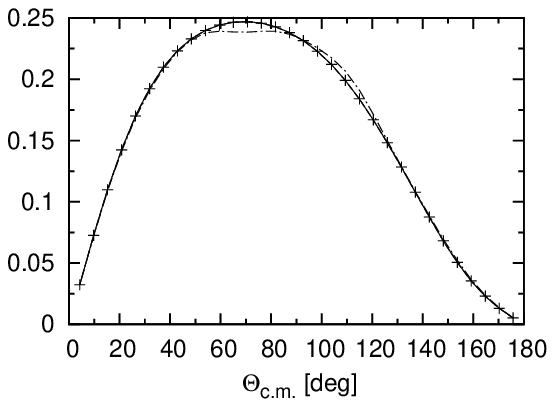,width=5cm,angle=0}
\epsfig{file=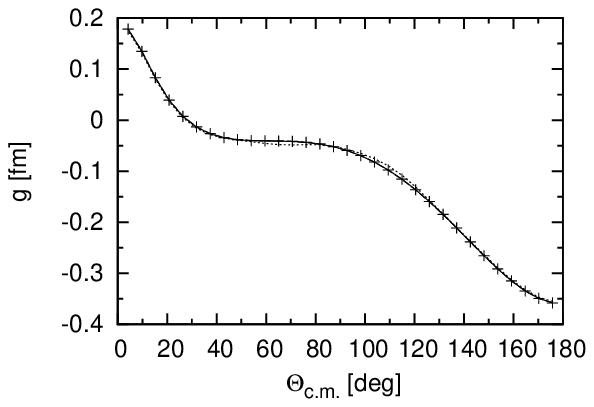,width=5cm,angle=0}
\epsfig{file=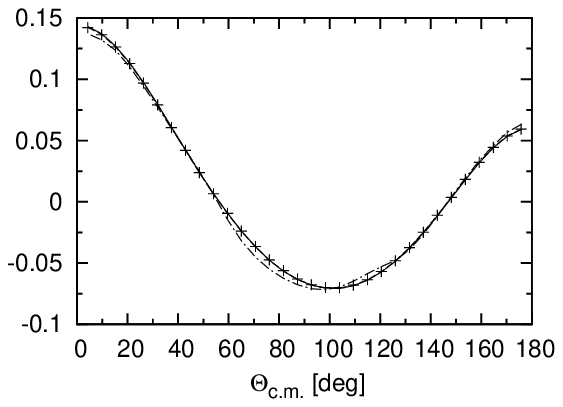,width=5cm,angle=0}
\epsfig{file=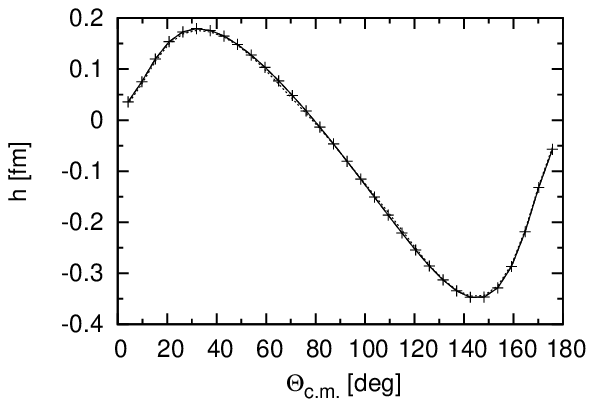,width=5cm,angle=0}
\epsfig{file=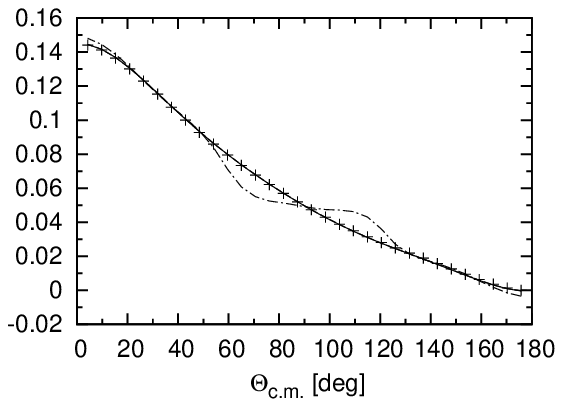,width=5cm,angle=0}
\epsfig{file=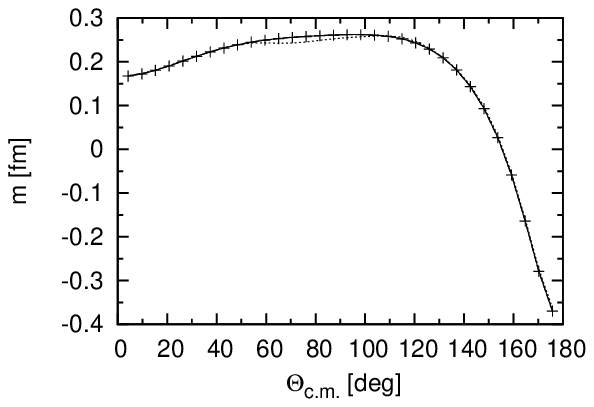,width=5cm,angle=0}
\epsfig{file=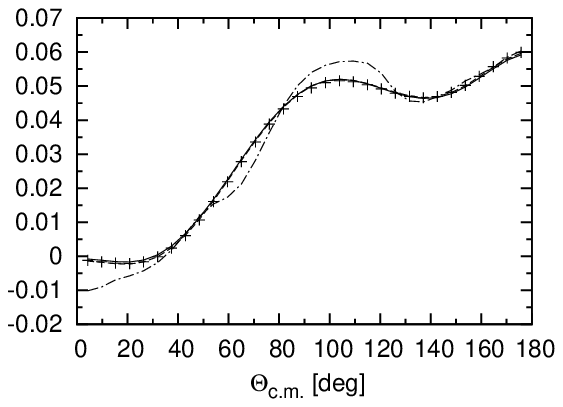,width=5cm,angle=0}
\caption{The Wolfenstein parameters for neutron-proton scattering for projectile
laboratory kinetic energy 300~MeV calculated with 
the chiral NNLO potential~\cite{evgeny.report}. 
The left panels show the real parts of the amplitudes, whereas the
imaginary parts are displayed in the right panels.
Crosses and lines show results of the same types of calculations 
as in Fig.~\ref{f1}.}
\label{f3}
\end{figure}

\section{Summary and conclusions}
\label{section4}

In \cite{2N3DIM} we formulated an
approach to the NN system employing spin-momentum operators 
multiplied by scalar functions, which only depend on the
momentum vectors. This representation of the NN potential leads to 
a system of six coupled equations for the scalar functions defining 
the NN t-matrix. This system of equations was first solved using the matrix
inversion method.

In the present paper we formulate two other methods to calculate 
the nucleon-nucleon t-matrix in the same 3D formulation of Ref.~\cite{2N3DIM}.
Our second method is iterative and uses a variant of the Lanczos algorithm 
\cite{stadler}.
In addition, as a third possibility, we treat the equation for the k-matrix
and obtain the on-shell t-matrix coefficients, solving an
additional equation, which connects
the on-shell scalar coefficients of the k- and t-matrices.
In the second step care is required, because only five spin-momentum operators
are independent and not all choices are numerically safe.
We show a very good agreement among all the considered methods 
for the selected neutron-neutron and neutron-proton observables. 
In these calculations we use the neutron-proton version 
of a chiral NNLO potential. 
In the case of the iteration method 
we demonstrate also a fast convergence with respect to the number of iterations.
This iterative method is best suited to be used in the 3N bound 
state and in the 3N continuum
calculations. In fact it has been already successfully used 
in the 3N bound state calculations~\cite{OUR3NBS3D}, 
realizing the theoretical formulation given in~\cite{2N3N}.

Last not least, we apply our 3D framework to the problem of proton-proton
scattering, using the proton-proton version of the chiral NNLO NN
potential, supplemented with the (screened) Coulomb force. We show converged
results for two different screening functions and find a good agreement with 
two other methods considered in \cite{romek}.


\section*{Acknowledgments}
We thank Dr. Evgeny Epelbaum for providing us with a code for the operator 
form of the chiral neutron-proton and proton-proton NNLO potentials.

We acknowledge support by the Foundation for Polish Science - MPD program,
co-financed by the European Union within the Regional Development Fund.

This work was supported by the Polish National Science Center
under Grant No. DEC-2011/01/B/ST2/00578.
It was also partially supported by
the European Community-Research Infrastructure Integrating Activity
``Exciting Physics Of Strong Interactions'' (acronym WP4 EPOS)
under the Seventh Framework Programme of EU.
The numerical calculations were partly performed on the
supercomputer cluster of the JSC, J\"ulich, Germany.


\end{document}